\documentclass[aps,twocolumn,showpacs,prl]{revtex4}
\usepackage[dvips]{graphicx}
\usepackage[english]{babel}
\usepackage[T1]{fontenc}
\usepackage{mathrsfs}
\usepackage[tbtags]{amsmath}
\usepackage{amssymb}
\usepackage{amsxtra}
\usepackage{amsopn}
\usepackage{latexsym}
\usepackage[mathcal]{eucal}
\usepackage{mathtools}

\newcommand{\BE}{\begin{equation}}
\newcommand{\EE}{\end{equation}}
\newcommand{\BA}{\begin{align}}
\newcommand{\EA}{\end{align}}

\newcommand{\nn}{\nonumber}

\renewcommand{\Re}{\mathop{\rm Re}}
\renewcommand{\Im}{\mathop{\rm Im}}

\begin{document}

\title{Dispersion relations for unphysical particles}

\author{Fabio Siringo}

\affiliation{Dipartimento di Fisica e Astronomia 
dell'Universit\`a di Catania,\\ 
INFN Sezione di Catania,
Via S.Sofia 64, I-95123 Catania, Italy}

\date{\today}

\begin{abstract}
Generalized dispersion relations are discussed for unphysical particles, e.g. confined degrees of freedom 
that are not present in the physical spectra but can give rise to observable bound states. 
While in general the propagator of the unphysical particles can have complex poles and cannot be reconstructed
from the knowledge of the imaginary part, under reasonable assumptions the missing piece of information is
shown to be in the rational function that contains the poles and must be added to the integral representation. 
For pure Yang-Mills theory, the rational part and the spectral term are identified
in the explicit analytical expressions provided by the massive expansion of the gluon propagator.
The multi particle spectral term turns out to be very small and the simple rational part provides, 
from first principles, an approximate propagator that is equivalent to the tree-level result of 
simple phenomenological models like the refined Gribov-Zwanziger model.
\end{abstract}





\maketitle

In many interacting theories and, notably, in non-Abelian gauge theories, some of the quantum fields 
describe {\it confined} particles that are not present in real spectra. They can be regarded as internal 
degrees of freedom of the theory and we can call them {\it unphysical} particles in this note, 
even when they play a very important role in the description of the phenomenology. For instance, gluons and quarks
are believed to be confined but their color singlet bound states are observed in physical spectra. More generally,
we are interested in the {\it physical} class of unphysical particles that give rise 
to observable bound states\cite{iparticle,dudalballs,capriballs}.

The propagators of these unphysical particles are usually studied in the Euclidean space where the correlators
emerge by 
lattice simulations\cite{bogolubsky,twoloop,dudal,binosi12,burgio15,bowman04,bowman05,su2glu,su2gho,duarte} 
or by numerical solution of a coupled set of 
integral equations\cite{aguilar10,alkofer,huber14,huber15g,huber15b,reinhardt14,gep2,varqed,varqcd,genself}. 
They are well
interpolated by regular functions on the {\it negative} real axis of the squared momentum $p^2=-p^2_E$.
However, their analytic continuation to the complex plane $z=p^2$ is not a trivial task at all,
because of the ill-defined problem of continuing a limited set of data points\cite{dudal14}.
Moreover, if the particle does not appear in the spectra, the general K\"allen-Lehmann representation does not
hold and, in the most studied case of QCD, the numerical data were shown to be not compatible with the standard
positivity constraints of the spectral functions. While that is usually regarded as an indirect evidence for
confinement, the same argument can question the whole existence of a spectral representation and the meaning
of the spectral functions. Actually, on general grounds, the usual dispersion relation between real and imaginary
part of the propagator does not hold for a generic unphysical particle, making even more ardue any guess of the
analytic properties in Minkowski space.

On the other hand, under general physical assumptions, an extension of the standard dispersion relation can be
proven, with a rational part that replaces the discrete spectral term of the physical particles.
While the proof is a trivial consequence of Cauchy formula, the generalized dispersion relation might be useful
for the study of physical bound states. Several model propagators, like that emerging from the refined 
Gribov-Zwanziger\cite{RGZ} and replica\cite{replica} models, 
are immediately shown to be equivalent to the most general propagator at
tree level by just neglecting the continuos term and retaining the rational part. Their phenomenological
success arises from the small weight of the continuos multi-particle spectral integral, enforcing the idea
that higher-order effects are small and can be treated by perturbation theory\cite{ptqcd0,ptqcd,ptqcd2,analyt}.

Let us assume that a generally analytic function $G(z)$ does exist, real and regular on the negative real 
axis $p^2=-p_E^2<0$ where $\Im G(-p_E^2)=0$ and that $G(-p^2_E)$ reproduces the numerical 
data points for the propagator in the Euclidean space.
If we knew that function exactly, then we could continue it analytically into the complex plane and 
reach the physical real axis $z=p^2>0$ where the imaginary function must have a branch cut unless $\Im G=0$. 
In fact, being real on the negative real axis, the function $G(z)$ must satisfy
\BE
G(z)^\star=G(z^\star)
\label{reality}
\EE
so that, in the limit $\epsilon\to 0$, for any $p^2>0$
\begin{align}
\Re G(p^2+i\epsilon)&=\Re G(p^2-i\epsilon)\nn\\
\Im G(p^2+i\epsilon)&=-\Im G(p^2-i\epsilon)
\label{RIG}
\end{align}
and the imaginary part must be discontinuous across the real axis unless it is exactly zero.
On general grounds, if $\Im G(p^2\pm i\epsilon)=0$ below a generic threshold $p^2<\theta$, then $G(z)$ must
have a branch cut along the positive real axis for $p^2>\theta\geq 0$.

Of course we do not have any further information on the analytic properties of the function $G(z)$ that,
in principle, might have any sort of singular points and branch cuts in the complex plane, out of the real axis.
However, we can somehow limit the arbitrariness by resctricting to the special class of particles with physical
bound states emerging in real spectra.
Bound states of unphysical particles arise in the spectral function of two-point correlators of
composite operators. These correlators can be built\cite{iparticle,dudalballs,capriballs}
by Feynman graphs in terms of the elementary
propagators $G(z)$ of the unphysical particles, thus inheriting the effects of their singularities.
Now, the physical two-point correlators that describe observable bound states must satisfy the general
K\"allen-Lehmann dispersion relations and must be analytic out of the real axis, with a single branch cut along
the positive real axis. That would pose a limit to the singularities of $G(z)$, even if in general, is quite
difficult to tell which kind of singularities are still acceptable.
For instance, it has been shown\cite{iparticle} that even when G(z) has a pair of
complex poles at $z=m^2_{\pm}=\alpha\pm i\beta$, the poles might combine in the composite correlator giving a 
physical branch cut along the real axis and a real multi-particle threshold at 
\BE
\theta^\prime=(m_++m_-)^2=2\left[\alpha+\sqrt{\alpha^2+\beta^2}\right]
\EE
as if they were real mass poles. Thus we cannot exclude the presence of simple poles everywhere in
the complex plane. 
The existence of other branch cuts in the compex plane, except for the
real axis, would be much more problematic.  We can hardly believe that entire branch cuts can totally
disappear in the correlator of the composite operators, without leaving any unphysical feature.
Thus from now on we will assume (or desire) that the most general propagator $G(z)$ of an unphysical particle has
no branch cuts except for $p^2>\theta\geq 0$ and might have any number of simple poles in the complex plane.
Because of Eq.(\ref{reality}), out of the real axis the poles can only occur in pairs of complex conjugated points,
while single poles might occur on the real axis below the threshold $\theta$. Moreover, we assume by physical
arguments that the function $G(z)$ goes to zero fast enough in the limit $z\to \infty$, say at least like $1/z$,
as satisfied by all the numerical data in the Euclidean space.

\begin{figure}[t] 
\centering
\includegraphics[width=0.3\textwidth,angle=-90]{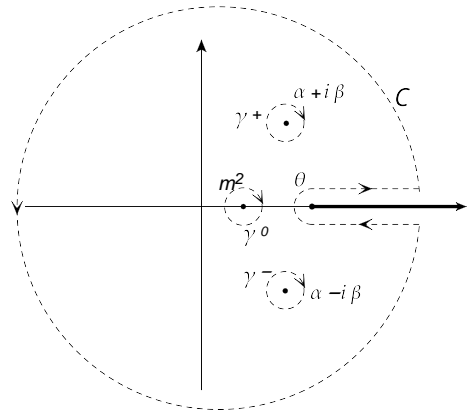}
\caption{Contour integration of Eq.(\ref{cauchy}) in the complex plane of the variable $\omega$. 
Just three of the simple poles are shown in the figure for the sake of clarity.}
\label{fig1}
\end{figure}

While reasonable, the hypotheses are strong enough to determine a generalized dispersion relation.
By Cauchy formula, the function $G(z)$ can be written as
\BE
G(z)=\frac{1}{2\pi i}\oint _\Gamma \frac{G(\omega)}{\omega- z}{\rm d}\omega
\label{cauchy}
\EE
where the contour $\Gamma=C \cup \gamma_i$ is shown in Fig.\ref{fig1} and $z$ is any complex point
inside the contour, where $G(z)$ is analytic. The single contours $\gamma_i$ are around the simple poles 
of $G(z)$ that occur at the complex conjugated points $z_i=m^2_{\pm}=\alpha_i\pm i\beta_i$ 
and on the positive real axis at $z_i=m^2>0$ 
(just three of them are displayed in Fig.{\ref{fig1}}).

The asymptotic behavior ensures that $G(\omega)/(z-\omega)\sim 1/\omega^2$ in the limit $\omega\to \infty$
so that sending the radius to infinity we can neglect the contribution of the external circle and, 
evaluating the single integrals over $\gamma_i$ by the residue theorem, we can write Eq.(\ref{cauchy}) as
\BE
G(z)=\sum_i\frac {R_i}{z-z_i}+\int_\theta^{+\infty} \frac{ \rho_c(\omega)}{z-\omega} {\rm d \omega}
\label{disp}
\EE
where $R_i$ is the residue of $G(z)$ in $z_i$ and $\rho_c(\omega)$ is the ususal continuos part of the 
spectral function
\BE
\rho_c(\omega)=-\frac{1}{\pi} \Im G(\omega+i\epsilon),\qquad \omega>\theta
\label{spectral}
\EE
having made use of Eq.(\ref{RIG}) in the integral above and below the branch cut and having set
$\epsilon\to 0$.
Eq.(\ref{disp}) holds exactly for any $z$ in the complex plane and generalizes the usual dispersion
relation in presence of an arbitrary number of simple poles. The main difference is in the rational
term that sums up the contributions of the poles.

In the very special case of a single pole on the real axis, the standard dispersion relation is recovered
by Eq.(\ref{disp}). In fact, when only a single real pole at $z=m^2<\theta$ contributes to the rational
part, Eq.(\ref{disp}) yields
\begin{align}
\Re G(p^2+i\epsilon)&=\frac{R}{p^2-m^2}+PV\int_\theta^{+\infty}\frac{\rho_c(\omega)}{p^2-\omega} \>{\rm d} \omega\nn\\
\Im G(p^2+i\epsilon)&=-\pi R \>\delta(p^2-m^2)-\pi \rho_c(p^2)
\label{RIG2}
\end{align}
and introducing a total spectral function $\rho (\omega)$
\BE
\rho(\omega)=-\frac{1}{\pi} \Im G(\omega+i\epsilon)=R \>\delta(\omega-m^2)+\rho_c(\omega)
\label{rho}
\EE
we can write
\BE
\Re G(p^2)=PV\int_0^{+\infty}\frac{\rho(\omega)}{p^2-\omega}\> {\rm d} \omega
\label{standard}
\EE
which is the standard dispersion relation between real and imaginary part of $G$ on the real axis.

Next, let us explore the case of a single pair of complex conjugated poles at $z=m^2_\pm=\alpha\pm i \beta$
and no poles on the real axis. That is a case encountered in several phenomenological models like the
refined version of the Gribov-Zwanziger\cite{RGZ} model and the replica model\cite{replica}.
Two complex conjugated poles also emerge from first principles by a massive expansion\cite{ptqcd,ptqcd2,analyt}
in Yang-Mills theory and seem to be a relevant model-independent physical feature of the gluon propagator.

Because of the reality condition Eq.(\ref{reality}), the residues must be complex conjugated and the rational
part in Eq.(\ref{disp}) reads
\BE
G^R (z)=\frac{R}{z-\alpha- i\beta}+\frac{R^\star}{z-\alpha+i\beta}.
\label{rational}
\EE
It is analytic on the real axis where its imaginary part $\Im G^R$ is zero. An important consequence is
that on the real axis the imaginary part of $G(z)$ does not know anything about the rational part and the
complex conjugated poles. In other words, at variance with the standard case of a single real pole in 
Eq.(\ref{standard}), {\it the real part cannot be reconstructed from the imaginary part} because no discrete
term arises in the total spectral function. Thus, the rational part $G^R$ must be added to the integral
of the continuos spectral function along the real axis
\BE
\Re G(p^2)= G^R(p^2)+PV\int_0^{+\infty}\frac{\rho(\omega)}{p^2-\omega}\> {\rm d} \omega
\label{ccpair}
\EE
where $\rho(\omega)=\rho_c (\omega)$ contains only the continuous term, as defined in Eq.(\ref{spectral}), 
while the function $G^R(p^2)$
is real and reads
\BE
G^R(p^2)=2\left[\frac{(p^2-\alpha)(\Re R)-\beta(\Im R)}{(p^2-\alpha)^2+\beta^2}\right].
\label{realrat}
\EE
From a mere formal point of view, we could recast Eq.(\ref{ccpair}) and make it look 
like the standard dispersion relation 
Eq.(\ref{standard}) by introducing a subtracted function $\delta G=(G-G^R)$, without poles, 
that would be entirely determined
by its imaginary part 
\BE
\Re \delta G(p^2)=PV\int_0^{+\infty}\frac{\rho(\omega)}{p^2-\omega}\> {\rm d} \omega
\label{dG}
\EE
where $\rho(\omega)$ is the same continuous spectral function of Eq.(\ref{spectral})
because $G^R(p^2)$ is real
\BE
\rho(\omega)=-\frac{1}{\pi} \Im \delta G(\omega+i\epsilon)=\rho_c (\omega).
\EE

However the subtracted function $\delta G$  would give only a part of the whole propagator and it turns out
to be a very small part, arising from the multi-patrticle continuous spectral function.

We may argue that, if the continuous contribution $\delta G$ is small, the whole propagator could be approximated by
the rational part $G(p^2)\approx G^R (p^2)$ that would capture the main physical properties. That would be
equivalent to the usual approximation $G(p^2)\approx R/(p^2-m^2)$ for a physical particle, when only the discrete
part of the spectral function is retained in Eq.(\ref{standard}).  In the Euclidean space we would get
the approximate tree-level propagator
\BE
G(-p_E^2)\approx G^R(-p_E^2)=-N \left[\frac{p_E^2+(\alpha+t\beta)}{p_E^4+2\alpha p_E^2+(\alpha^2+\beta^2)}\right].
\label{ratE}
\EE
where $N$ is a renormalization constant and $t=(\Im R)/(\Re R)=\tan[\arg(R)]$. That would just be a
re-parametrization of the refined Gribov-Zwanziger propagator\cite{RGZ} with the parameters 
\begin{align}
M^2_{GZ}&=\alpha+t\beta\nn\\
m^2_{GZ}&=\alpha-t\beta\nn\\
\lambda^4_{GZ}&=\alpha^2+\beta^2.
\label{GZpar}
\end{align}
Thus, the propagator arising from that model can be seen as the most general tree-level
propagator with a pair of complex conjugated poles and its phenomenological success seems to be related 
to the relatively small weight of the continuos term in the general dispersion relation of Eq.(\ref{disp}).

It is instructive to check the dispersion relation of Eq.(\ref{ccpair}) and the tree-level approximation
of Eq.(\ref{ratE}) by a direct calculation for pure Yang-Mills theory where explicit analytical
expressions are available for the propagators at one-loop and third order 
of the massive expansion\cite{ptqcd,ptqcd2,analyt} in Landau gauge. The one-loop gluon propagator can be continued
to Minkowski space and studied in the complex plane\cite{analyt}. The optimized expansion is in remarkable agreement
with the data of lattice simulations in the Euclidean space and predicts the existence of just two complex conjugated
poles for the gluon propagator, with  no poles on the  real axis. We use the explicit expression of Ref.\cite{ptqcd2}
for the gluon propagator $\Delta(p)$ optimized for $SU(3)$ and analytically continued to the complex plane, 
as discussed in Ref.\cite{analyt}.
The poles are found at the complex masses $m^2_\pm=(\alpha,\pm \beta)=(0.163,\pm 0.602)$ GeV$^2$
with $t=\pm 3.17$. 

\begin{figure}[t] 
\centering
\includegraphics[width=0.35\textwidth,angle=-90]{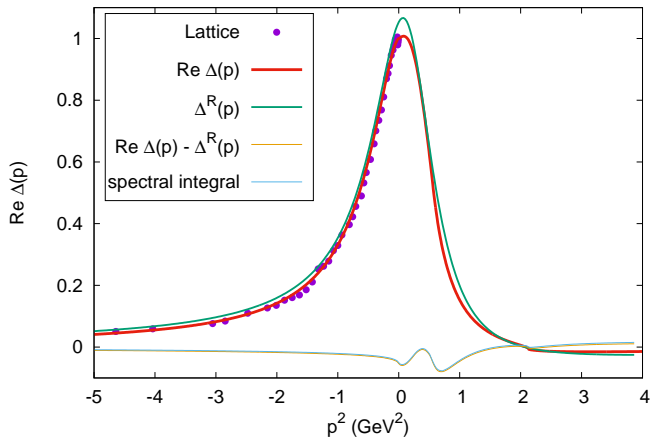}
\caption{The real part of the one-loop gluon propagator $\Delta(p)$ (red line) is
evaluated for $SU(3)$ pure Yang-Mills theory by the optimized massive expansion of Ref.\cite{ptqcd2} and 
analytically continued to Minkowski space as discussed in Ref.\cite{analyt}.  
The lattice data of Ref.\cite{bogolubsky} are shown for comparison.
The green line is the rational part $\Delta^R$ evaluated by Eq.(\ref{realrat}).
At the bottom, the subtracted function $\delta \Delta=\Re \Delta-\Delta^R$ is reported twice,as arising from
the exact difference of the analytical expressions and from the numerical integral of the spectral
function in Eq.(\ref{dG}). The two curves are not distinguishable at the
scale of the figure.}
\label{fig2}
\end{figure}

\begin{figure}[t] 
\centering
\includegraphics[width=0.35\textwidth,angle=-90]{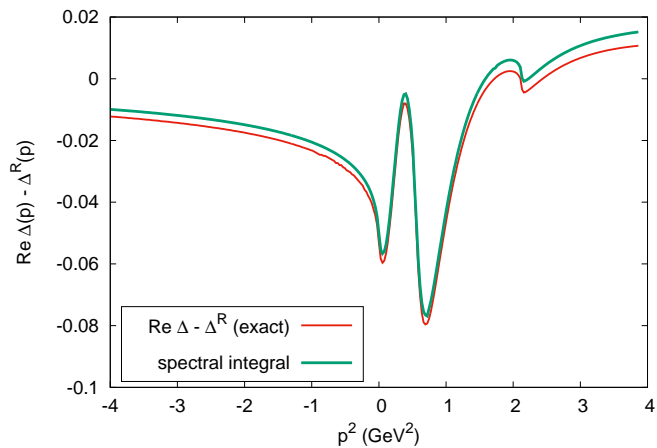}
\caption{The same subtracted function $\delta \Delta=\Re \Delta-\Delta^R$ of Fig.\ref{fig2} is shown in
more detail by a very enlarged scale. The red line is the exact difference between the analytical expressions.
The green line is the numerical reconstruction of the same function by the spectral integral in Eq.(\ref{dG}).
The slight difference arises because of the numerical accuracy and of the finite cut-off in the integral.}
\label{fig3}
\end{figure}

The real part of
the optimized one-loop gluon propagator $\Delta(p)$ is shown in Fig.\ref{fig2} together with the lattice 
data of Ref.\cite{bogolubsky}. In the same figure, the rational part $\Delta^R(p)$ is also shown, evaluated 
according to Eq.(\ref{realrat}). We observe that the approximation of Eq.(\ref{ratE}) 
is very well satisfied over the whole range
of positive and negative $p^2$. The difference between the curves is equal to the subtracted 
function $\Re \Delta-\Delta^R$
that is shown at the bottom of Fig.\ref{fig2}. The subtracted function is first evaluated directly, 
subtracting the analytical expressions. Then, the same subtracted function is calculated numerically 
by integration of the continuous spectral function in Eq.(\ref{dG}). The two results are not distinguishable at the
scale of Fig.\ref{fig2}. Their comparison is shown in more detail at a very large scale in Fig.\ref{fig3}  where
a small difference arises because of the numerical accuracy and of a finite cut-off in the integration.

Even if the subtracted function has a rich behaviour that arises from the multiparticle continuous 
spectral function, its absolute value is very small and
the total propagator seems to be very well described by the simple rational part. Higher-order corrections 
are probably still smaller, enforcing the idea that the one-loop propagators 
of the massive expansion\cite{ptqcd,ptqcd2,analyt}
provide a very good approximation of the exact propagators in the whole complex plane.

In summary, Eq.(\ref{disp}) generalizes the ususal dispersion relation to the case of an unphysical particle with
complex masses. The general integral representation of Eq.(\ref{disp}) can be useful as a building block
for the evaluation of the correlators of composite operators. For instance, the standard scheme of 
dimensional regularization can be used by combining the denominators. In general, the real part of the 
propagator cannot be reconstructed by the knowledge of the imaginary part that does not contain information
on the pairs of complex poles. Standard disperison relations can be written for the subtracted propagators, 
once the diverging rational part has been subtracted. However, the main physical content seems to be in the
rational part that provides a tree-level approximation for the propagator and is entirely determined by the
complex masses and the residues, thus generalizing the renormalized propagator of an interacting physical particle.

For pure Yang-Mills theory, the generalized dispersion relation in Eq.(\ref{disp}) allows us to separate the
multi-particle contribution arising from the spectral function and the rational tree-level term that contains
the poles. An explicit one-loop calculation shows that the gluon propagator is very well described by the
rational part and the spectral function adds only a small correction. Thus, from first principles, a tree-level
model propagator arises that captures the physics of the gluon sector and is equivalent to the
refined Gribov-Zwanziger model\cite{RGZ}.
It is remarkable that the model propagator is just the most general tree-level propagator with a pair of
complex conjugated poles and its phenomenological success is a direct consequence of the small
weight of the spectral function in the gluon sector.
Moreover, here the model propagator arises in a more general context, without any special assumption about Gribov
copies or the existence of condensates. Without any phenomenological assumption, the approximate rational
propagator in Eq.(\ref{ratE}) emerges as the natural tree-level approximation whenever the exact propagator
has a pair of complex conjugated poles and the multiparticle continuous contribution of the spectral function
is small. Actually, from first principles, by a massive expansion for the exact Faddeev-Popov Lagrangian in
Landau gauge, those requirements have been shown to be fully satisfied by the one-loop gluon propagator.
Then, the tree-level propagator is fully determined by the complex masses and the residues that can be easily
extracted from the analytical one-loop expressions, without any phenomenological parameter. Once determined,
that simple propagator could be useful 
for the study of bound states and glueball masses\cite{iparticle,dudalballs,capriballs}.

Finally, having shown that the continuos spectral function gives a very small contribution 
to the gluon propagator, higher-order terms are expected to be even smaller, thus
enforcing the idea that the massive expansion of Ref.\cite{ptqcd,ptqcd2,analyt} could provide a very
reliable analytical tool for the study of QCD in the infrared.

\end{document}